\documentclass[a4paper,12pt]{article}

\usepackage{amsmath}
\usepackage{graphicx}
\usepackage{hyperref}
\usepackage{color}
\usepackage{subfigure}
\usepackage{cite}
\usepackage{array}
\usepackage{textcomp}

\def\beq{\begin{equation}}
\def\eeq{\end{equation}}
\def\barr{\begin{array}}
\def\earr{\end{array}}

\begin{document}
\begin{center} 
\begin{large}
\textbf{Bottom Pair Production and 
Search for Heavy Resonances}
\end{large}\\
\vspace*{20pt} 
Pratishruti Saha\\
\vspace*{5pt}
\begin{footnotesize}
Department of Physics and Astrophysics, 
University of Delhi, Delhi 110007, India.
\end{footnotesize}\\
\end{center} 

\vspace*{40pt}

\begin{abstract} 
\noindent 
The search for heavy resonances has for long been a part
of the physics programme at colliders.
Traditionally, the dijet channel has been examined as part of 
this search.
Here, $b\bar b$ production is examined as a possible search channel.
The chiral color model (flavor universal as well as non-universal)
and the flavor universal coloron model are
chosen as templates of models that predict the existence of heavy
colored gauge bosons.
It is seen that, apart from the resonance, 
the interference of the Standard Model and new physics amplitudes
could provide a useful signal.
Of particular interest, is the case of the 
non-universal chiral color model, as this channel may allow the model
to be confirmed or ruled out as the reason behind the forward-backward
asymmetry in $t \bar t$ production.

\vspace*{40pt}
\noindent
\texttt{PACS Nos:14.70.Pw,14.65.Fy,12.40.-y} \\ 
\texttt{Key Words:axigluon,coloron,Tevatron,LHC,$b\bar b$ etc.} 
\end{abstract}

\vspace*{40pt}

\section{Introduction}
\label{sec:intro}

The Standard Model (SM) seeks to describe Nature as a realization
of the gauge group $SU(3)_C \otimes SU(2)_L \otimes U(1)_Y$. 
While there exists substantial experimental evidence to suggest
that this is indeed correct, at least up to the scale of a few 
hundred GeVs, the picture is far from complete. 
Several extensions of the SM have been suggested~\cite{NP_reviews} 
and continue to be suggested in the 
attempt to redress the `unsatisfactory' aspects of the model.
One common feature among many of these models is the 
existence of massive particles that couple to a pair of
SM fermions and are likely to appear as a resonance in the
process $f \bar f \to f' \bar f'$.
Experimental searches for such particles are most often carried out
in the dijet channel or in the Drell-Yan process.
However, as one wishes to study a fermion-antifermion final state,
the $b \bar b$ channel is also an option that could be investigated.
If the new particles under consideration have only strong interactions,
then the Drell-Yan process would not be sensitive to their presence.
As for the dijet process, while it may receive contributions from
new strongly interacting particles, sensitivity would be limited by 
the fact that final states consisting of a quark-antiquark pair
not be distinguishable from those with $qq$, $\bar q \bar q$, $qg$, 
$\bar qg$ or $gg$. On the other hand, $b$-jets can be identified with
reasonable accuracy using flavor-tagging techniques.
Thus, $b \bar b$ production may prove to be useful as a search channel.

In this paper, the reach of the $b \bar b$ channel in the
search for some classes of new physics(NP) models, namely,
the chiral color model (with and without flavor universality)
and the flavor universal coloron model, is examined.
This channel is of particular importance for the
flavor non-universal chiral color model. The observation of 
forward-backward asymmetry in $t \bar t$ production($A_{FB}^t$)
caused a slew of models to be proposed as 
plausible explanations. In a majority of these, new couplings were 
introduced for the top quark while keeping bottom quark couplings unchanged.
The nu-axigluon is an exception to this and a search in the $b \bar b$ channel 
can provide one way to distinguish this model amongst a host of others.

The next section contains a brief description of the models 
and the existing limits on their constituents.
The details of the calculation are discussed in 
Sections~\ref{sec:analytic} and \ref{sec:numeric}.

\section{Models}
\label{sec:models}

In the Standard Model, the gauge group $SU(2)_L \otimes U(1)_Y$
is broken to $U(1)_{em}$. 
This has prompted attempts to examine whether QCD may be the remnant
of a broken symmetry too. The unifiable chiral color model and the 
flavor universal coloron model are two models which propose that 
$SU(3)_C$ is actually a relic of a $SU(3) \otimes SU(3)$ symmetry 
broken spontaneously at a high scale.

Chiral color models~\cite{ChiralColor_1} assume the gauge group
describing strong, weak and electromagnetic interactions to be 
$SU(3)_L \otimes SU(3)_R \otimes SU(2)_L \otimes U(1)_Y$. 
$SU(3)_{R-L}$ is sought to be broken spontaneously at a scale
comparable to the scale of electroweak symmetry breaking. 
$SU(3)_{R+L}$ remains and is identified with $SU(3)_C$.
Thus, in these models, there exists an octet of massive colored gauge
bosons (axigluons) alongside an octet of massless ones (gluons).
The axigluons($A$) have an axial vector coupling to quarks which has the 
\textit{same} strength ($g_s$) as the gluon-quark coupling.
These models also require the existence of additional fermions and
colored scalars.
Infact, in the most optimistic scenario~\cite{ChiralColor_2},
five generations of quarks and leptons, three Higgs doublets 
and additional electrically neutral as well as charged fermion
multiplets in `non-standard' representations are predicted.
This gives rise to a model that, besides replicating many of
the successes of the Standard Model, is rich in high scale physics
and is unifiable at a scale much lower than that for the latter.

Initially the scale of chiral-color breaking was assumed to be
the same as that of electroweak symmetry breaking and axigluons
were expected to have mass $\sim$ 250 GeV.
Early experimental bounds obtained from measurements of $\Upsilon$
decays and hadronic cross-sections in 
$e^+ e^-$ collisions~\cite{limits_e+e-} ruled out $M_A <$ 50 GeV.
The region 50 GeV $< M_A <$ 120 GeV was ruled out by 
considering effects on hadronic decays of $Z^0$ and the possibility of associated production of axigluons~\cite{Doncheski:1998ny}.
Dijet production in hadronic colliders has been repeatedly surveyed for signals of a resonant axigluon~\cite{Bagger:1987fz}.
A series of searches at the Tevatron in this 
channel~\cite{dijet_older} have now resulted in exclusion of 
$M_A <$ 1250 GeV at 95\% confidence~\cite{dijet_2008}.
The use of forward-backward asymmetry
\footnote{Axial-vector coupling of axigluons to quarks implies that interference between gluon-mediated and axigluon-mediated processes
can give rise to a forward-backward asymmetry. 
This is discussed in detail later.}
as a signal for axigluons has also been studied~\cite{Sehgal:1987wi} 
and possible limits from top production data have been considered in 
Refs.~\cite{limits_top,DC_and_RMG,Melnitchouk:2008ij}.

More recently, flavor non-universal versions of the original 
chiral color model have been proposed~\cite{Frampton:2009rk,Ferrario:2009bz}
as possible explanations of the 
forward-backward asymmetry observed in $t \bar t$ production
at the Tevatron~\cite{CDF_asymm,CDF_asymm_latest}.
In particular, the model in Ref.~\cite{Frampton:2009rk}
contains four quark generations and is based on the gauge group
$SU(3)_A \otimes SU(3)_B \otimes SU(2)_L \otimes U(1)_Y$.
The gluon and the flavor non-universal axigluon($A'$)
are admixtures of the gauge bosons corresponding to 
$SU(3)_A$ and $SU(3)_B$ with $\theta_{A'}$ being the 
mixing angle.
The coupling of the non-universal axigluon~\footnote{This will 
henceforth be referred to as the nu-axigluon for purposes 
of disambiguation.} 
consists of a vector and an axial-vector part.
While the vector coupling is generation universal 
($- g_s \, \cot2\theta_{A'}$), the axial-vector coupling is not, with 
$g_A^q = - g_s \,{\rm cosec} 2\theta_{A'}$ for the first two generations 
and $g_A^t = + g_s \,{\rm cosec} 2\theta_{A'}$ for the other two.
Demanding that the couplings be perturbative, restricts 
10\textdegree $< \theta_{A'} <$ 45\textdegree.

Although, the Lorentz structure of the couplings is 
the similar to that in the original chiral color model, 
the non-universal nature of the couplings implies that
the mass limits on the former from the dijet search,
are not directly applicable.
However, as the main motivation behind the proposition was to
explain the observed $A_{FB}^t$,
the parameter space can be constrained using measurements
in the top sector, such as the $t \bar t$ cross-section, 
$A_{FB}^t$ and the $m_{t \bar t}$ spectrum~\cite{Frampton:2009rk}.
In particular, the apparent agreement of the invariant mass
distribution (which is reported for $m_{t \bar t}$ upto 1400 GeV)
with the SM, can be used immediately, albeit somewhat naively,
to put a lower limit of 1400 GeV on $M_{A'}$.

In the flavor universal coloron model~\cite{flavor_universal_coloron},
the high scale color gauge group is \mbox{$SU(3)_I \otimes SU(3)_{II}$}. 
This is broken to $SU(3)_C$ at the TeV scale. Here again, there is an 
octet of massive colored gauge bosons (colorons) in addition to gluons. 
The original model~\cite{Hill_1991} was aimed at constructing a 
dynamical mechanism for electroweak symmetry breaking involving a 
$\langle\bar tt\rangle$ condensate. 
In this model, the third generation quarks belonged to a different 
representation of $SU(3)_I \otimes SU(3)_{II}$ as compared to the other
quark families. However, in the flavor universal version of the model, 
all quarks transform as $(1,3)$ under the extended color gauge group. 
The couplings are proportional to $\xi_1$ and $\xi_2$ for $SU(3)_I$ and 
$SU(3)_{II}$ respectively with $\xi_1 \ll \xi_2$. The coupling of the 
coloron($C$) to quarks is then proportional to $\gamma_\mu\cot\xi$, 
where, $\xi$ is the mixing angle and $\cot\xi$ = $\xi_2/\xi_1$. 
An additional scalar multiplet, transforming as $(3,\bar 3)$, 
effects the symmetry breaking. Initially, this model was proposed 
in order to explain excess seen in the inclusive jet cross-section 
in the high $E_T$ region by the CDF experiment at the Tevatron~\cite{dijet_excess}.
With increase in statistics and improvement in both theoretical 
calculations and exprimental techniques, the agreement between 
theory and experiment has improved considerably~\cite{dijet_excess_gone}.
However, the model itself continues to be of interest as it can
accomodate, within its framework, a theory with composite quarks~\cite{flavor_universal_coloron}. 
Further, as in the case of the original topcolor proposal~\cite{Hill_1991,Hill_1995},
the flavor universal version too can provide a scheme for dynamical
EWSB via formation of a $\langle\bar tt\rangle$ condensate~\cite{Popovic:1998vb}.

The original proponents of the model~\cite{flavor_universal_coloron},
placed the limit $M_C/\cot\xi >$ 450 GeV required to keep corrections 
to the electroweak $\rho$ parameter within allowed limits~\cite{rho_limit}.
In addition, demanding that the model remain in its Higgs phase at low energies, results in an upper limit 
$\sim$ 4 on the value of $\cot\xi$~\cite{Simmons:1996fz}.
The phenomenology of colorons was studied in detail in Ref.~\cite{Simmons:1996fz,Bertram:1998wf}
wherein dijet data from the Tevatron~\cite{Abe:1995jz,Abbott:1998wh,Abbott:1998yy} 
was used to place a lower limit of 870 GeV and 1 TeV on $M_C$ for
$\cot\xi$ values of 1 and 2 respectively, and the lower limit on 
$M_C/\cot\xi$ was raised to 837 GeV.
Sensitivity to this variety of new physics is also expected in 
the top sector and this has been explored in Refs.~\cite{Hill_and_Parke,DC_and_RMG}. 
The latest measurement of dijet mass spectrum at the CDF experiment 
at the Tevatron, however, rules out the existence of flavor-universal colorons
with mass below 1250 GeV~\cite{dijet_2008}.

\subsection{Search Efforts}
As mentioned earlier, in the search for axigluons and colorons,
the dijet channel has been studied extensively and has been
the focus of most experimental searches.
Rates have been calculated for on-shell production of 
axigluons/colorons followed by decay and this has been used for
comparison with data.
Some searches have also been carried out 
in the $t\bar t$ channel~\cite{Melnitchouk:2008ij}.
It is clear that a (nu-)axigluon/coloron resonance, 
if present, will also affect $b\bar b$ production rates. 
While both the CDF and D0 experiments have vast B-physics
programmes, they are mostly concerned with studying properties of
B-mesons~\cite{Tevatron_B-Physics}.
The potential of the $b\bar b$ channel in searches for heavy 
resonances remains largely untapped.

In the case of the models described above, $qg$ and $gg$ dijet 
final states are not sensitive to the new particles and 
create a background. On the other hand, t-channel processes
such as $qq' \rightarrow qq'$, while getting contributions
from new physics, tend to render difficult, the task of 
identification of a resonance structure in the dijet invariant 
mass spectrum. This is specially true when $M_{boson} \sim$ 1 TeV 
and the resonance is a broad one to begin with~\footnote{Efficiency 
factors associated with the reconstruction of jets also lead to
broadening of the resonance peak. However, for (nu-)axigluons and 
colorons in the mass range$~\sim$ 1 TeV, the natural width itself
is large.}.
On the other hand contribution to $b \bar b$ production from the
$t$-channel is negligible. This, coupled with advancements in 
$b$-tagging algorithms may be exploited in strengthening the search for 
\mbox{(nu-)axigluons} and colorons as well as other new particles
with similar interactions.

\section{$b\bar b$ production}
\label{sec:analytic}

At a hadron collider, $b\bar b$ production gets contributions from the 
processes $q\bar q \rightarrow b\bar b$ and $gg \rightarrow b\bar b$.
At the centre-of-mass energies associated with currently operational
colliders, namely, the Tevatron and the LHC, 
production is dominated by the gluon initated process.
However, in the high invariant mass region (in which we are interested),
it is the quark initiated process which dominates.

The presence of (nu-)axigluons or colorons modifies the amplitude
for the quark initiated process.
The $b$ density in protons and anti-protons is negligible and hence 
the major contribution accrues from the s-channel process
$q\bar q \to b\bar b$, mediated by a (nu-)axigluon or a coloron 
in addition to a gluon\footnote{Electroweak contributions can be 
neglected as they are suppressed by a factor 
$\alpha_{EW}^2/\alpha_s^2$ and hence small.}.
This makes for a distinct, albeit broad, 
peak\footnote{provided the mass is within the c.m. energy range
of the collider} in the $b\bar b$ invariant mass ($m_{b\bar b}$) spectrum.
The analytic expressions for the differential cross-section
are analogous to those for ${t \bar t}$ 
production~\cite{DC_and_RMG,Frampton:2009rk}
with $m_t \to m_b \approx 0$. 
The width for the new gauge bosons is about 10\% of the mass. 
Thus, for $M_{new}\sim$ a few hundred GeVs, the large width implies 
that the narrow-width approximation is no longer valid and 
the off-shell contribution must also be taken into account. 
For axigluons and nu-axigluons, there are terms proportional 
to odd powers of $\cos\theta$ which give rise to a 
forward-backward asymmetry in the angular distribution.
In contrast, $\cos\theta$ dependence for the coloron case
is identical to that in the pure SM and is forward-backward symmetric.
The interference between the coloron and gluon mediated
amplitudes is negative in the region $\hat s < M_C^2$
and causes the $m_{b\bar b}$ spectrum to dip before peaking.

The gluon initiated process remains unaffected by the presence of 
(nu-)axigluons/colorons and forms the chief SM background. 
The corresponding analytic expression is available in 
Ref.~\cite{Combridge:1978kx}. Here, the $t$-channel and $u$-channel 
contributions get enhanced in the region where 
$\cos\theta\rightarrow$1, i.e. in the low $p_T$ region.
Moreover, the low threshold for $b\bar b$ production implies that
it is easily attained with low values of Bjorken $x$, for which, 
gluon densities are larger than quark densities. 
Hence, the dominant contribution, particularly
in the low $p_T$ and low $\sqrt{\hat s}$ region, arises from this process.

\section{Numerical Results}
\label{sec:numeric}

In this study, the contribution of (nu-)axigluons and colorons
to $b\bar b$ production is calculated at the parton level. 
CTEQ6L parton distribution functions~\cite{CTEQ} 
are used.
The factorization scale is chosen to be $E_T$.
The renormalization scale for $\alpha_s$ is $E_T$ everywhere 
except in the calculation of decay-widths, where, it is set to be
the mass of the relevant boson.
Other chosen parameters include $\alpha_s(M_Z)$ = 0.118 
(consistent with CTEQ6L), \mbox{$m_t$ = 172 GeV}~\cite{PDG}
and all other \mbox{$m_q$ = 0}.
Production rates are computed for the Tevatron ($\sqrt{s}$=1.96 TeV)
as well as the LHC ($\sqrt{s}$=7 TeV). 

As mentioned earlier, there is a an enhanced contribution to 
the SM $gg \rightarrow b\bar b$ process from the 
low $p_T$ and $\sqrt{\hat s}$ regions.
To reduce this background, appropriate cuts need to be imposed on 
$p_T$. Further, the identification of a $b \bar b$ final state 
requires a double b-tag. b-tagging efficiency is low for 
high rapidity($y$) regions~\cite{CDF_btag}
and this restricts the $y$-range that can be
taken into account.
The details of the choice of cuts, efficiency factors etc. 
for the two colliders are given in Table~\ref{tab:cuts}. 

\begin{table}[!htbp]
\begin{center}
\begin{tabular}{| c | c | c | c | c |}
\hline
~ & $\epsilon_b$ & $\epsilon_{mistag}$ & $p_T^{min}$ & $|y|^{max}$ \\
\hline
TeV &  \texttt{0.3} & \texttt{0.030} & \texttt{100 GeV} & \texttt{1.0} \\
\hline
LHC &  \texttt{0.4} & \texttt{0.012} & \texttt{500 GeV} & \texttt{1.3} \\
\hline
\end{tabular}
\end{center}
\caption{\em $p_T^{min}$ is the value of $p_T$
at which $q \bar q \to b \bar b$ starts dominating over $gg \to b \bar b$.
The values mentioned above correspond to the minimal choice
and lead to maximum signal significance. 
The $|y|$ cuts are designed to exclude the regions where b-tagging efficiency 
is very \mbox{low~\cite{CDF_btag,ATLAS_btag}}.}
\label{tab:cuts}
\end{table}

\begin{figure}[!htbp]
\centering
\includegraphics{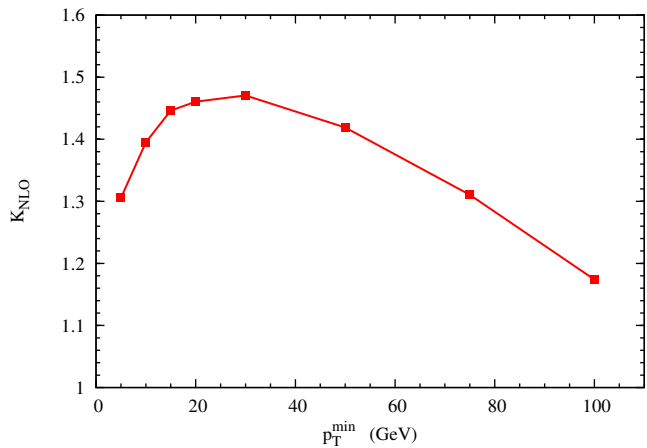}
\caption{\em Dependence of the NLO K-Factor on the $p_T$ cut.}
\label{fig:KFactors}
\end{figure}

Contributions to the cross-section also appear from the 
next-to-leading order(NLO).
The new particles being heavy, the additional contribution
to the total cross-section from the NLO is expected to be 
dominated by the SM corrections and hence, in the absence of 
full NLO calculations incorporating contributions 
from the new physics models under consideration,
only the SM K-Factor for $b \bar b$ production at the Tevatron
is calculated. 
MC@NLO~\cite{MCatNLO} is used for this purpose.
The dependence of the K-Factor on the $p_T$ cut is also studied.
For the range of $p_T^{min}$ examined, it is seen~\ref{fig:KFactors} that,
the dependence is mild and the K-Factor varies between 1.17 and 1.47.

\subsection{At the Tevatron}

\begin{figure}[!htbp]
\centering
\subfigure[]
{
\includegraphics[width=3.1in,height=2.5in]{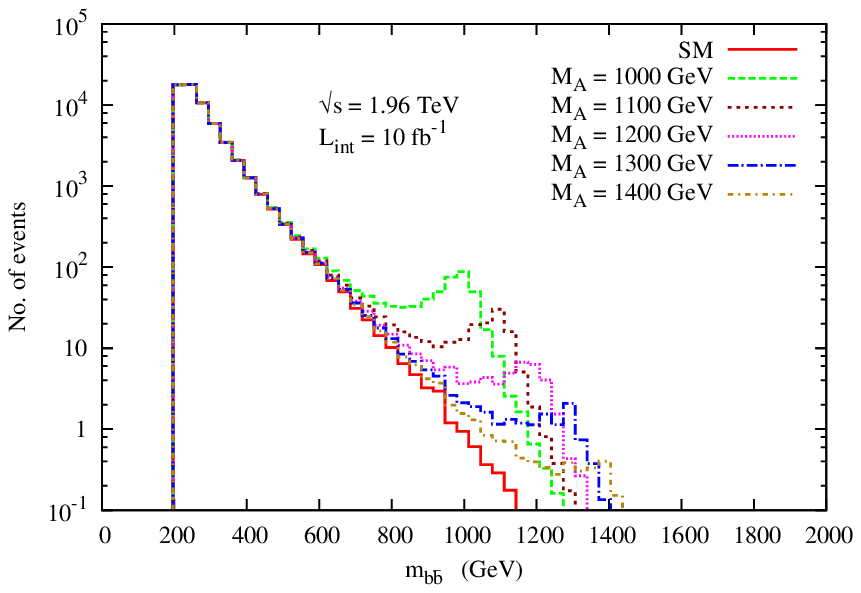}
\label{fig:axi_mbb_TeV}
}
\subfigure[]
{
\includegraphics[width=3.1in,height=2.5in]{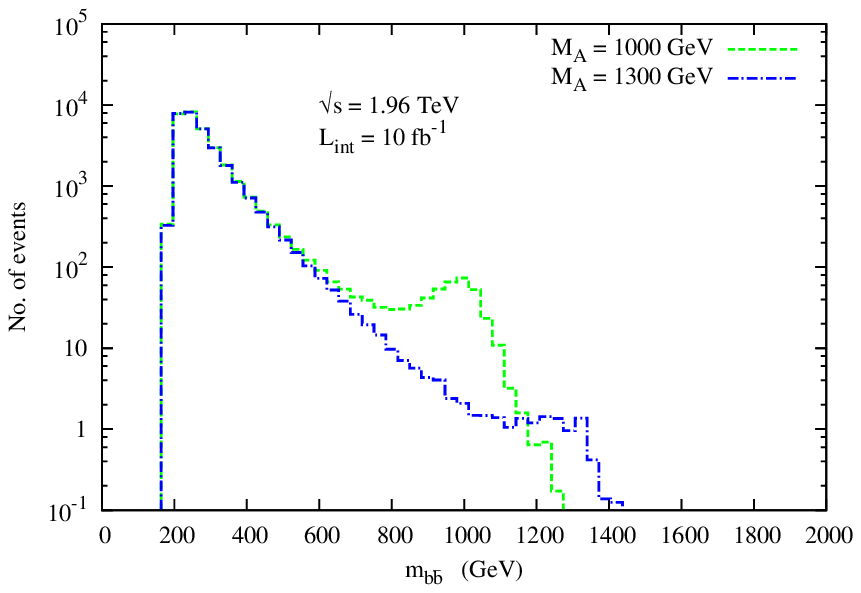}
\label{fig:axi_mbb_TeV_smeared}
}
\caption{\em $b \bar b$ invariant mass spectrum at the Tevatron in 
the presence of axigluons. {\em (b)} shows the effective broadening
due to jet reconstruction.}
\label{fig:axi_invmass_TeV}
\end{figure}

Fig.~\ref{fig:axi_mbb_TeV} shows the invariant mass spectrum of the 
$b\bar b$ pair in the presence of axigluons at the Tevatron
for an assumed integrated luminosity of 10 fb$^{-1}$.
Note that apart from the SM $b \bar b$ production, the background 
also gets a contribution from the SM dijet process due to possible
mis-identification of light jets. 
In addition, bottom pairs are produced
in $t \bar t$ events with almost 100\% efficiency.

\begin{figure}[!htbp]
\centering
\subfigure[]
{
\includegraphics[width=3.1in,height=2.5in]{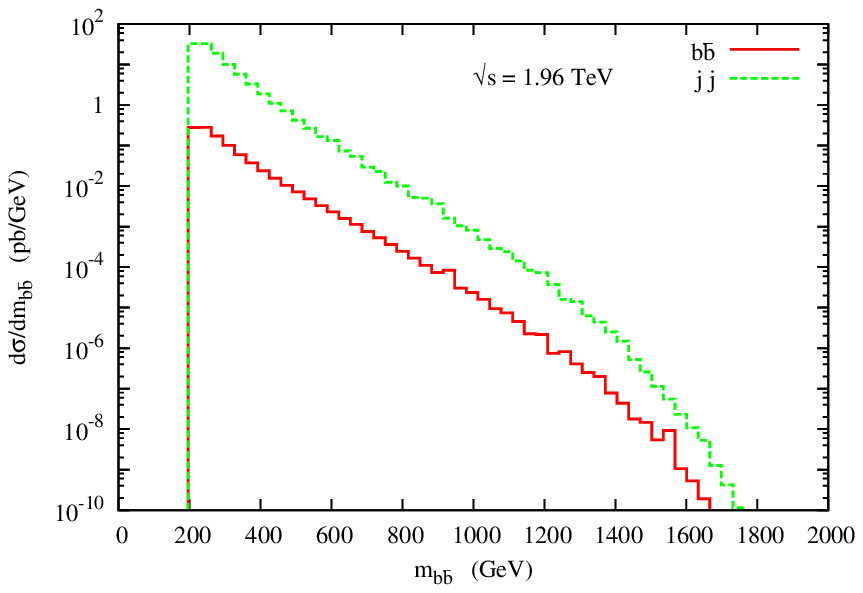}
\label{fig:SM_bg_noeff}
}
\subfigure[]
{
\includegraphics[width=3.1in,height=2.5in]{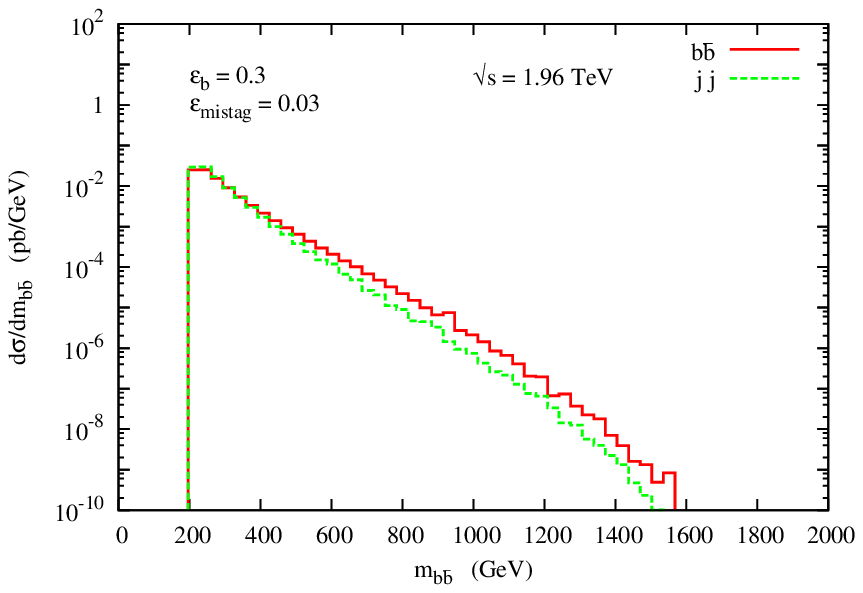}
\label{fig:SM_bg_witheff}
}
\caption{\em Comparison between the SM backgrounds at the Tevatron 
due to $gg \to b \bar b$ and due to mistagging of dijets.}
\label{fig:SM_background}
\end{figure}

While, the $t \bar t$ background may be eliminated by demanding
that there are no additional hard jets or isolated hard leptons
associated with the event, there is no such straightforward 
scheme to do away with the dijet background and this must be 
taken into consideration while calculating signal significance.
The invariant mass spectra for the SM $b \bar b$ and dijet 
processes are compared in Fig.~\ref{fig:SM_background}
which shows that, once the respective tagging and mistagging
probabilities are taken are taken into account,
the dijet background plays only a subdominant role.
Nevertheless, this contribution has been included in the
distributions shown here. The NLO K-Factors are 1.17~\ref{fig:KFactors}
and 1.3~\cite{dijet_2008,dijet_NLO} for $b \bar b$ and dijets, respectively.

Returning to Fig.~\ref{fig:axi_mbb_TeV}, one sees that
a resonance peak is clearly observable above the net
SM background for $M_A$ upto 1300 GeV. Even for higher 
masses ($\sim$ 1400 GeV), a deviation in the tail of the 
distribution seems apparent although this region is plagued
by low statistics.
In the experimental scenario, it is expected that, 
the sharpness of any existent resonance peak would be worsened 
to some extent due to detector resolution effects and errors 
associated with the reconstruction of jets.
In order to estimate the influence of such effects, the energy
of the outgoing jets is smeared with a Gaussian distribution 
whose variance is given by the energy resolution of the 
central hadron calorimeter
($\sigma_{E_T}/E_T$ = 50\%/$\sqrt{E_T({\rm GeV})} \oplus$ 3\%)~\cite{dijet_2008}.
The effect of the smearing is shown in 
Fig.~\ref{fig:axi_mbb_TeV_smeared} for two representative cases of 
\mbox{$\rm M_A$=1000 GeV} and \mbox{$\rm M_A$=1300 GeV}.
While a broad resonance is still distinguishable for the former,
in the latter case, though the excess in the tail is conspicuous, the
identification of a resonance structure appears somewhat difficult.

\begin{figure}[!htbp]
\centering
\includegraphics[width=3.1in,height=2.5in]{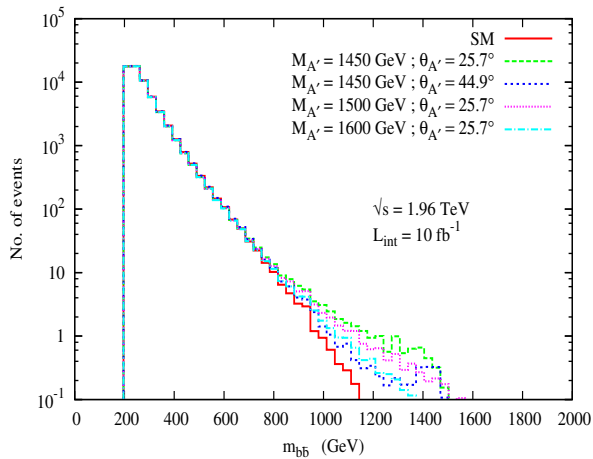}
\caption{\em $b \bar b$ invariant mass spectrum at the Tevatron 
in the presence of flavor non-universal axigluons. Parameters have been
chosen so that $\sigma_{t \bar t}$~\cite{CDF_csec} and 
$A_{FB}^t$~\cite{CDF_asymm_latest} measurements are respected 
at the 1-$\sigma$ level. $M_{A'} >$ {\em 1400 GeV} to be consistent 
with the $m_{t \bar t}$ spectrum~\cite{CDF_mtt}.}
\label{fig:newaxi_invmass_TeV}
\end{figure}

Fortunately, the presence of a resonance in the invariant mass spectrum
need not be the sole indicator of the existence of axigluons.
At the Tevatron, it will also be signalled by 
forward-backward asymmetry($A_{FB}^b$) 
in $b \bar b$ production\footnote{At the LHC, the initial state is
symmetric and no \textit{simple} forward-backward asymmetry w.r.t the 
beam direction can be defined, although possible ways of constructing 
analogous observables that will probe the same effect have been discussed
in Refs.~\cite{Rodrigo-98,Rodrigo-99,asymm_LHC_rest}}.
The value of $A_{FB}^b$ can be calculated using various observables.
For a given observable ${\cal O}$, $A_{FB}^b$ is defined as

\[
 A_{FB}^b = \frac{ \sigma({\cal O}>0) - \sigma({\cal O}<0) }
               { \sigma({\cal O}>0) + \sigma({\cal O}<0) } \ .
\]

The cosine of the angle made by the outgoing bottom quark with the direction 
of the proton beam ($\cos\theta_b$) and the difference in the rapidities
of the bottom and the anti-bottom ($\Delta y$) are two observables most 
often used in this context.
Of these, $\Delta y$ gives the value of $A_{FB}^b$ that would be 
measured in the centre-of-mass frame as it is invariant under boosts
in the longitudinal direction. $\cos\theta_b$, on the other hand, 
gives $A_{FB}^b$ in the laboratory frame. 

\begin{figure}[!htbp]
\centering
\subfigure[]
{
\includegraphics[width=3.1in,height=3.0in]{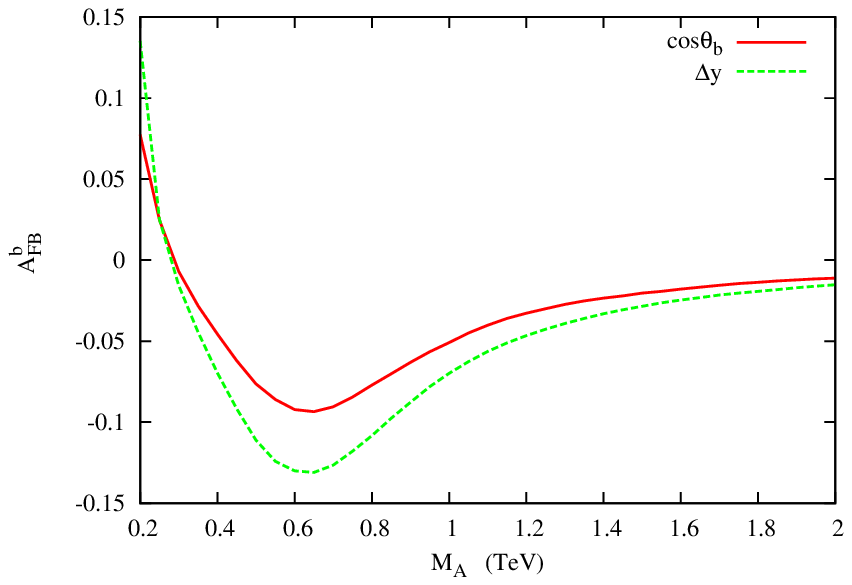}
\label{fig:axi_afb_full}
}
\subfigure[]
{
\includegraphics[width=3.1in,height=3.0in]{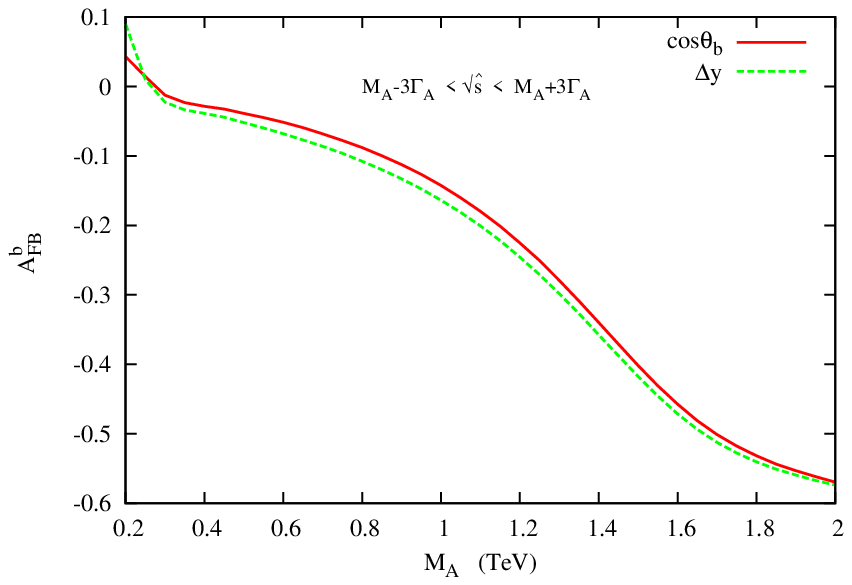}
\label{fig:axi_afb_inrange}
}
\caption{\em Variation in $A_{FB}^b$ with axigluon mass.}
\label{fig:axi_asymm}
\end{figure}

The variation in $A_{FB}$ with axigluon mass is seen in 
Fig.~\ref{fig:axi_asymm}.
In Fig.~\ref{fig:axi_afb_full} values of $A_{FB}^b$ as 
obtained
\footnote{In the figures presented here, only 
tree-level new physics contributions to $A_{FB}^b$ are depicted. 
There is also some contribution from the SM as discussed later
in the text.} using $\cos\theta_b$ as well as $\Delta y$ 
are plotted as a function of axigluon mass. 
For most of the $M_A$ range, negative asymmetries are predicted.
Note that, the asymmetry is expected to be more manifest in the
region of the phase space where the dominant contribution to 
the cross-section comes from the axigluon mediated sub-process.
This, clearly, is the region where $\sqrt{\hat s} \approx M_A$.
If the forward-backward asymmetry is calculated in a $\rm 3\Gamma_A$ 
interval around the resonance (Fig.~\ref{fig:axi_afb_inrange},
a monotonic behaviour is seen with the magnitude of the asymmetry 
growing with the mass of axigluon for $M_A >$ 400 GeV.

\begin{figure}[!htbp]
\centering
\subfigure[]
{
\includegraphics[width=3.1in,height=3.0in]{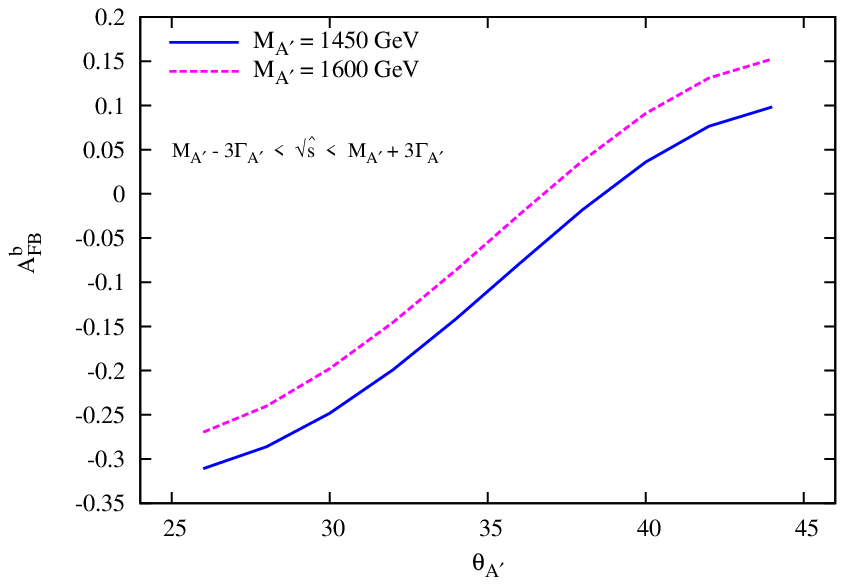}
\label{fig:newaxi_afb}
}
\subfigure[]
{
\includegraphics[width=3.1in,height=3.0in]{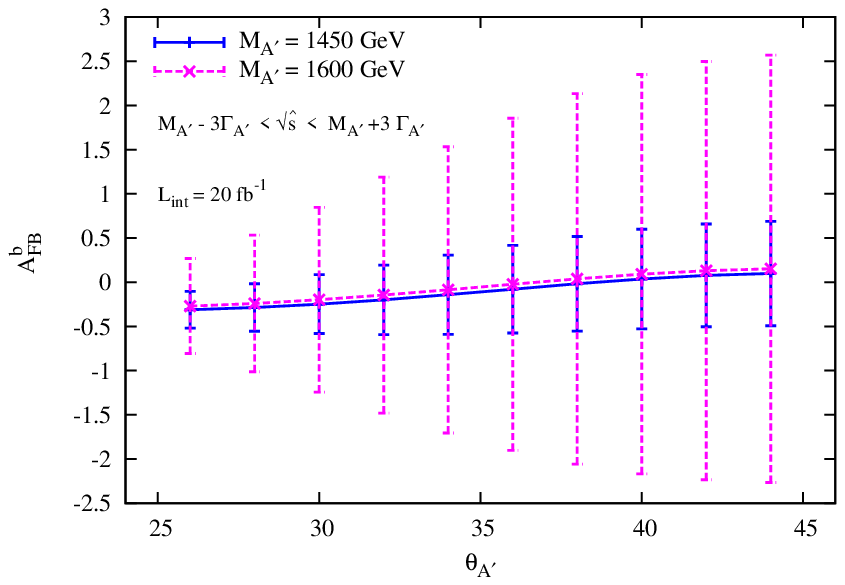}
\label{fig:newaxi_afb_werr}
}
\caption{\em Variation in $A_{FB}^b$ with coupling for nu-axigluon.{\em (b)} shows
an estimate of the statistical errors.}
\label{fig:newaxi_asymm}
\end{figure}

Contribution to $A_{FB}^b$ also comes from the SM electroweak 
production of $b \bar b$ pairs. However, the magnitude of the 
contribution (as in the case of cross-section) is small. 
Further, $A_{FB}^b$ of a few percent is expected due NLO QCD 
effects~\cite{Rodrigo-98,Rodrigo-99}.
Fig.3 of Ref.\cite{Rodrigo-98} (Fig.5 of Ref.~\cite{Rodrigo-99})
shows the expected asymmetry in 
$b \bar b$ production from NLO QCD as a function of $\sqrt{\hat s}$.
The asymmetry is positive and of the order of 5\%-6\% for 
350 $< \sqrt{\hat s} <$ 1800 GeV.
In comparison, consider Fig.~\ref{fig:axi_afb_inrange}
which shows $A_{FB}^b$, not as a function of $\sqrt{\hat s}$, 
but in a region where the value of $\sqrt{\hat s}$ lies close to $M_A$.
Since  $A_{FB}^b$ is a smooth and a slowly varying 
function of $M_A$ (equivalently,  $\sqrt{\hat s}$), this correspondence 
is quite accurate.
As can be seen, for significant parts of the parameter space, 
the asymmetries are, generically, large and negative. 
This also holds true for nu-axigluons as shown by 
Fig.~\ref{fig:newaxi_afb}. Thus, in both the cases, the behaviour of the 
new physics contribution to $A_{FB}^b$ is quite different from that of the SM 
contribution and tends to dominate the latter. 
Hence, the combined effect of SM and new physics tends to 
result in a significant negative value for the net asymmetry. 
In the event that such a negative asymmetry is observed, it would indicate 
contributions from such models.

The measurement of $A_{FB}^b$, however, depends strongly on the accuracy
with which the charge of the b-jet can be measured. 
Although such a measurement was reported at the LEP~\cite{LEP_bB_asymm}, 
the complex detector environment at a hadron collider makes this an even
more challenging task at the Tevatron.
A measurement of forward-backward asymmetry would be particularly
interesting in the case of nu-axigluons, where, such a measurement
would allow the model to be singled out as the cause for the asymmetry
in the top sector~\footnote{This was also pointed out recently in Ref.~\cite{Rizzo}.}
In Fig.~\ref{fig:newaxi_asymm}, $A_{FB}^b$ is plotted
as a function of $\theta_{A'}$. Large asymmetries are seen to be predicted.
But even a naive estimate (considering only statistical errors)
shows that the errors involved are large~\ref{fig:newaxi_afb_werr}. 
This is simply because the $\hat s$ region where 
contribution from new physics is maximum is close to the limit
of the energy reach of the Tevatron. Hence event rates are low
and error bars are large.

Fig.~\ref{fig:newaxi_invmass_TeV} shows the invariant mass distribution
for the case of the non-universal axigluon. Representative values of 
$M_{A'}$ and $\theta_{A'}$ are chosen from the parameter space allowed
by the $\sigma_{t \bar t}$~\cite{CDF_csec} and 
$A_{FB}^t$~\cite{CDF_asymm_latest} measurements at the 1-$\sigma$
level. $M_{A'}$ is restricted to above 1400 GeV in order to respect
constraints from the measured $m_{t \bar t}$~\cite{CDF_mtt} distribution.
Deviations above the background are clearly seen.
Note that larger values of $\theta_{A'}$ correspond to smaller couplings.

\begin{figure}[!htbp]
\centering
\subfigure[]
{
\includegraphics[width=3.1in,height=2.5in]{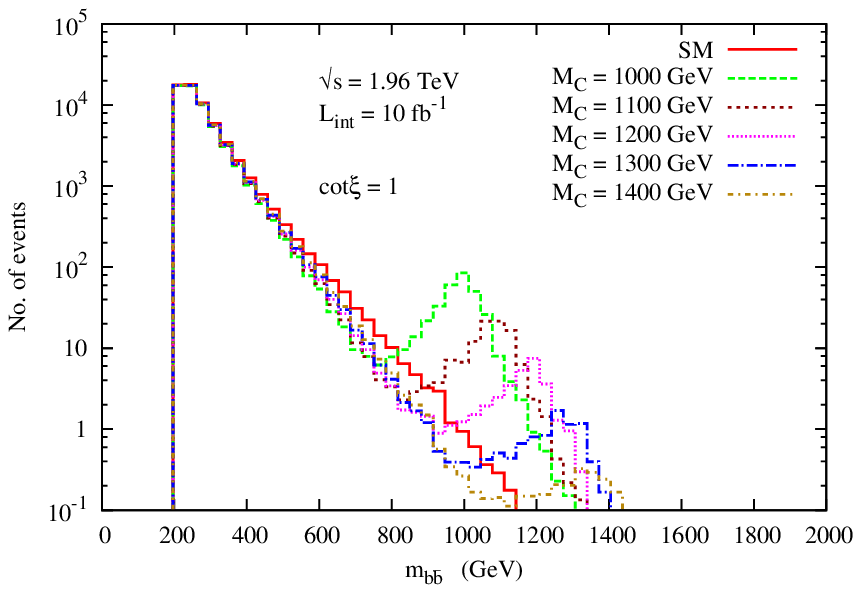}
\label{fig:col_1_mbb_TeV}
}
\subfigure[]
{
\includegraphics[width=3.1in,height=2.5in]{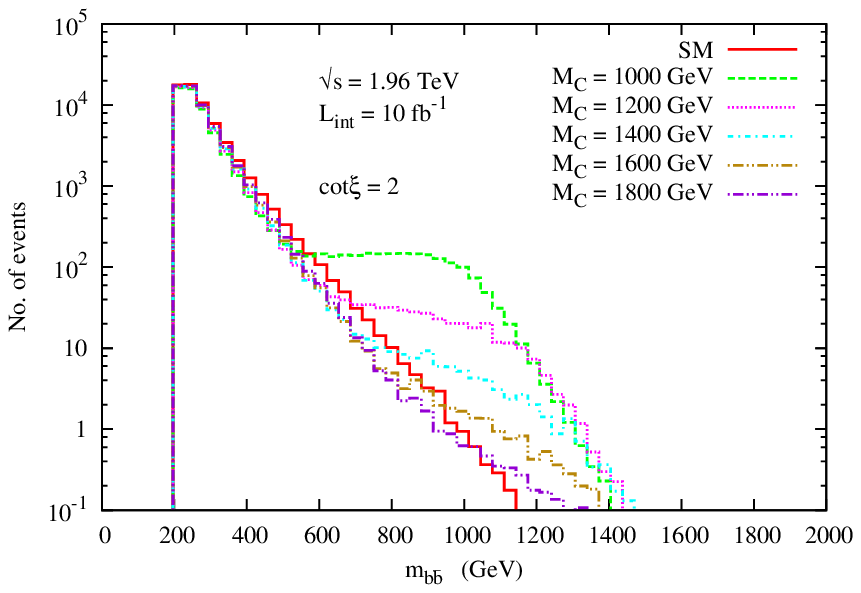}
\label{fig:col_2_mbb_TeV}
}
\caption{\em $b \bar b$ invariant mass spectrum at the Tevatron 
in the presence of colorons with 
{\em (a)}$\cot\xi$={\em 1} and 
{\em (b)}$\cot\xi$={\em 2}}
\label{fig:col_invmass_TeV}
\end{figure}

In the case of the coloron, the deviation in the 
$m_{b \bar b}$ spectrum is apparent even in the region 
much below the peak. 
The invariant mass distributions are plotted  
in Fig.~\ref{fig:col_invmass_TeV} 
taking $\cot\xi$=1 and $\cot\xi$=2 
as two representative cases for $M_C$ values in the 
range 1000 GeV to 1600 GeV, along with the Standard Model
background.
It is seen that the spectrum dips below the Standard Model
expectation before rising at the resonance. 
In the case $\cot\xi=1$, while the resonance would allow 
the identification of colorons of mass upto about even 1300 GeV, 
the suppression may signal the presence of colorons of mass 
upto 1600 GeV. This characteristic suppression of production rates in the 
low $m_{b\bar b}$ region can be used to attribute any excess 
present in the high $m_{b\bar b}$ region to a coloron, thus
distinguishing it from an axigluon.

For $\cot\xi$=2, the resonance is very broad.
This is likely to make the determination of coloron mass, 
a difficult task. Nevertheless, the spectrum is decidedly different
from what is expected in the Standard Model. 
An excess is clearly noticeable for $M_C$ upto 1400 GeV. 
Here too, the suppression proves to be more useful and may be used to 
detect colorons with $M_C$ upto \mbox{1800 GeV}.
Thus, the $b\bar b$ channel can be used to extend the 
search for colorons at the Tevatron beyond currently available
limits from the dijet channel.

\subsection{At the LHC}

At the LHC, the domination of the gluon initated process increases
even more, creating the requirement for a more stringent $p_T$ cut.
The signal suffers a further drop due to diminished the anti-quark fluxes
in a $pp$ collider. 

\begin{figure}[!htbp]
\centering
\subfigure[]
{
\includegraphics[width=3.1in,height=2.5in]{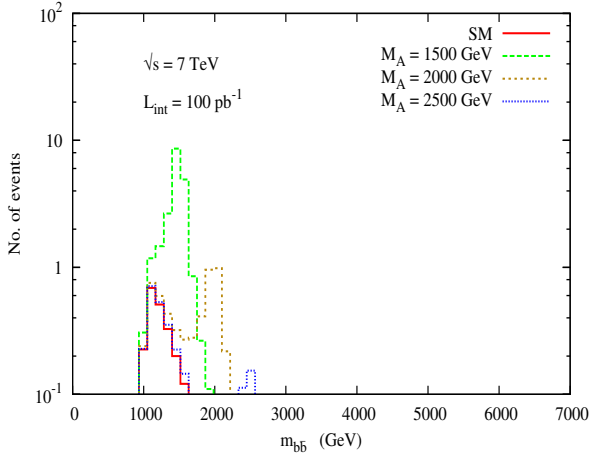}
\label{fig:axi_mbb_LHC}
}
\subfigure[]
{
\includegraphics[width=3.1in,height=2.5in]{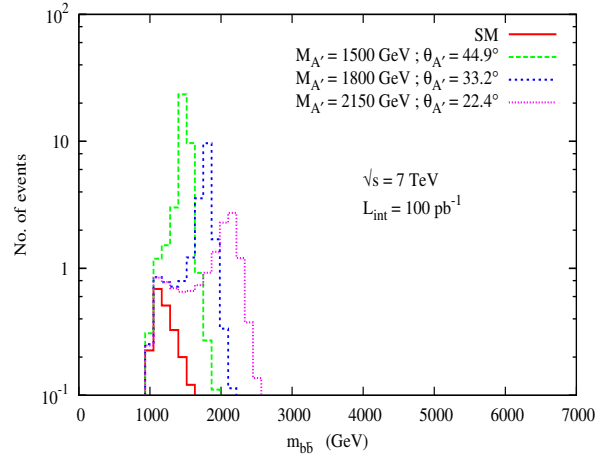}
\label{fig:new_mbb_LHC}
}
\caption{\em $b \bar b$ invariant mass spectrum at the LHC in 
the presence of {\em (a)}axigluons and {\em (b)}nu-axigluons.
The kinematic cuts mentioned in Table~\ref{tab:cuts} have been used.}
\label{fig:(nu-)axi_invmass_LHC}
\end{figure}

\begin{figure}[!htbp]
\centering
\subfigure[]
{
\includegraphics[width=3.1in,height=2.5in]{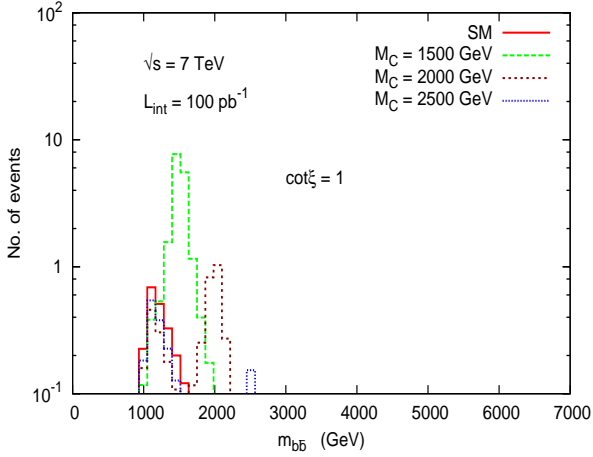}
\label{fig:col_1_mbb_LHC}
}
\subfigure[]
{
\includegraphics[width=3.1in,height=2.5in]{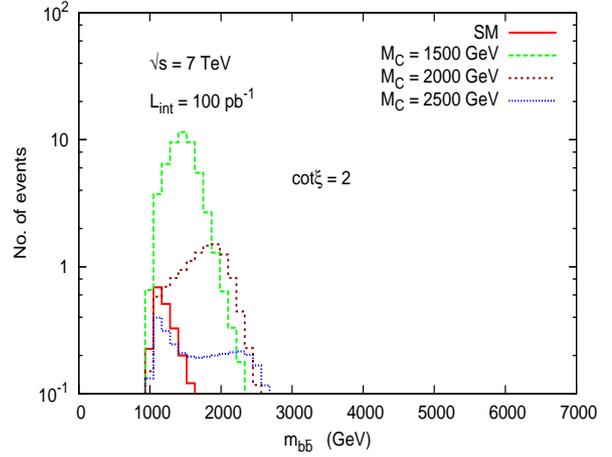}
\label{fig:col_2_mbb_LHC}
}
\caption{\em $b \bar b$ invariant mass spectrum at the LHC in 
the presence of colorons.
The kinematic cuts mentioned in Table~\ref{tab:cuts} have been used.}

\label{fig:col_invmass_LHC}
\end{figure}

However, inspite of this, greater centre-of-mass energy allows
the search to be extended into mass regions $\sim$ 2.2 TeV. 
The $m_{b \bar b}$ distributions for the 
different new physics scenarios are shown in 
Fig.~\ref{fig:(nu-)axi_invmass_LHC} and \ref{fig:col_invmass_LHC},
assuming $\sqrt{s}$ = 7 TeV and integrated luminosity 100 pb$^{-1}$.
Of course, finite detector resolution will cause a broadening 
of the peak, nevertheless, the deviation will be sufficient 
so as to be considered an unambiguous signal of new physics.

Apart from the $m_{b\bar b}$ spectrum, the $p_T$ spectrum also
gets modified in all of the above cases. The sensitivity of the
$p_T$ spectrum to new physics is similar to that of the 
invariant mass spectrum. 
However, the latter fares slightly better and hence the 
$p_T$ distributions are not presented here. 
For the case of the coloron, these have been considered 
in detail in Ref.~\cite{Trott}.
Of course, in the event
that new physics is observed, a correlated deviation in the $m_{b \bar b}$
and the $p_T$ spectrum would only serve to further strengthen
the claim.

\section{Summary}
\label{sec:summary}

The bottom pair production process at the Tevatron as well at the LHC
can be surveyed for signals of axigluons (flavor universal as well non-universal)
and colorons.
While all the classes of particles will appear as resonances in the $b\bar b$
invariant mass distribution, at the Tevatron, the measurement of a 
forward-backward asymmetry will be an additional indication of the 
existence of \mbox{(nu-)axigluons}.
On the other hand, deficient event rates in the low and intermediate $m_{b\bar b}$
regions will signal the presence of colorons.
While measurement of mass may be difficult for these are all broad resonances,
(particularly the coloron, when $\cot\xi > 1$), the deviation would be 
sufficient to warrant an explanation from physics beyond the 
Standard Model.
The $b \bar b$ channel can also be used to identify the non-universal axigluons
as the reason behind the intriguing observation of forward-backward asymmetry
in $t \bar t$ production.

\section*{}

\begin{center}
\begin{small}\textbf{ACKNOWLEDGEMENT}\end{small}\end{center}
PS would like to thank CSIR, India for assistance under 
SRF Grant 09/045(0736)/2008-EMR-I.



\end{document}